\documentclass[final,5p,times,twocolumn]{elsarticle}
\usepackage{amsmath,latexsym,amssymb,geometry}
\usepackage{setspace,stackrel,tikz,graphicx,caption}
\usepackage{exscale,relsize,subfig,textcomp,tikz,stackrel,setspace,float}
\usepackage{mdframed,pifont,empheq,algpseudocode,algorithm,MnSymbol}
\usepackage{silence,mwe}
\newcommand{\EE}{\mathrm{E}}

\begin{document}

\begin{frontmatter}	
	\title{Steepest Descent Multimodulus Algorithm\\ for Blind Signal Retrieval in QAM Systems}
	\author[Habib]{Shafayat~Abrar}
	\address[Habib]{Associate Professor\\ School of Science and Engineering\\ Habib University, Gulistan-e-Jauhar\\ Block 18, Karachi 75290, Pakistan\\ Email: \textrm{\texttt{shafayat.abrar@sse.habib.edu.pk}}}

\begin{abstract}
We present steepest descent (SD) implementation of multimodulus algorithm (\textsf{MMA2-2}) for blind signal retrieval in digital communication systems. In comparison to stochastic approximate (gradient descent) realization, the proposed SD implementation of \textsf{MMA2-2} equalizer mitigates inter-symbol interference with relatively smooth convergence and superior steady-state performance.      	

\end{abstract}
\end{frontmatter}


\section{Introduction}

The multimodulus algorithm (\textsf{MMA2-2})  \cite{abrar2010ieeetn,abrar2006ieeespl} is given as
\begin{eqnarray}\label{EqMMA22}
\boldsymbol{w}_{n+1} =
\boldsymbol{w}_n   + \mu\,\big(R_m-y_{R,n}^2\big)y_{R,n}\boldsymbol{x}_n-\textrm{j} \,\mu\,\big(R_m-y_{I,n}^2\big)y_{I,n}\boldsymbol{x}_n,
\end{eqnarray}
where, \(\textrm{j}=\sqrt{-1}\), $R_m$ is a positive statistical constant, $\boldsymbol{x}_n$ is channel observation vector, $\boldsymbol{w}_n$ is equalizer vector, and \(y_n=\boldsymbol{w}_n^H\boldsymbol{x}_n=y_{R,n}+\textrm{j}\, y_{I,n}\) is  equalizer output. The update (\ref{EqMMA22}) is probably the most popular and widely studied multimodulus algorithm capable of equalizing multi-path transmission channel blindly and recovering carrier phase jointly in quadrature amplitude modulation based wireless, wired and optical communication systems.
The update, however, is stochastic approximate in nature, works on symbol-by-symbol basis, and is relatively slower in convergence when compared to its batch counterparts. Moreover, even in successfully converged state, the error function in update expression is non-zero except for instances when \(|y_{\cdot,n}|=\sqrt{R}\); these fluctuations (as quantified in \cite{azim2015}) cause delay in switching to decision-directed mode and lead to decision errors causing loss of information.

In order to exploit full potential of \textsf{MMA2-2}, there is a new practice in literature to realize it in batch mode. In this context, Han \textit{et al.} discussed a number of methods including steepest descent implementation for constant modulus algorithms (\textsf{CMA}) and relaxed convex optimization for \textsf{MMA2-2} in \cite{han2012thesis} and \cite{han2017convex}, respectively. In \cite{shah2015}, Shah \textit{et al.} discussed batch \textsf{MMA2-2} by exploiting iterative blind source separation framework and came up with Givens and hyperbolic rotations based batch \textsf{MMA2-2}.
Also in \cite{daumont2010}, authors transformed \textsf{MMA2-2} cost into an analytical problem and solved that for both batch and adaptive processing using subspace tracking methods. The most rigorous treatment appeared in \cite{Debals2016}, where a batch \textsf{MMA2-2} is obtained which included an analytical transformation to a set
of coupled canonical polyadic decompositions by using subspace methods.
Recently, Han and Ding \cite{GiHong2012} suggested a steepest descent batch implementation of a class of \textsf{CM} algorithms where the update process did not require equalizer outputs (no feedback) and rather relied directly on statistics obtained from the received signal. Motivated by that approach, in this correspondence, we present a steepest descent implementation of \textsf{MMA2-2} by estimating required batch statistics iteratively while maintaining simplicity of its adaptive structure. To the best of our knowledge, a steepest descent implementation of \textsf{MMA2-2} has not been realized in literature.

\section{Feedforward Steepest Descent Algorithms}

In order to realize a steepest descent implementation of (\ref{EqMMA22}), we need to estimate expected value of its error function. 
\begin{equation}\label{EqCMAparray}
\boldsymbol{w}_{n+1} =
\boldsymbol{w}_n   + \mu\,\EE\Big[\big(R_m-y_{R,n}^2\big)y_{R,n}\boldsymbol{x}_n-\textrm{j}\big(R_m-y_{I,n}^2\big)y_{I,n}\boldsymbol{x}_n\Big]
\end{equation}
We evaluate this expectation in forward driving manner as advocated in \cite{GiHong2012}. According to which, we replace $y_n$ with $\boldsymbol{w}_n^H\boldsymbol{x}_n$, and evaluate statistical average of matrix quantities involving $\boldsymbol{x}_n$ conditioned on $\boldsymbol{w}_n$.
Exploiting the facts that
\begin{subequations}\label{EqFacts}\begin{alignat}{1}
y_{R,n}&=\textstyle\frac{1}{2}\Big({\boldsymbol{w}_n^H\boldsymbol{x}_n + \boldsymbol{x}_n^H\boldsymbol{w}_n}\Big)\\
y_{I,n}&=\textstyle\frac{1}{2\textrm{j}}\Big({\boldsymbol{w}_n^H\boldsymbol{x}_n - \boldsymbol{x}_n^H\boldsymbol{w}_n}\Big)
\end{alignat}
\end{subequations}
and after some manipulations, we obtain
\begin{equation}\begin{split}
&\EE\Big[\big(R_m-y_{R,n}^2\big)y_{R,n}\boldsymbol{x}_n-\textrm{j} \big(R_m-y_{I,n}^2\big)y_{I,n}\boldsymbol{x}_n\Big]\\
&~~=\EE\Big[\Big(R_m\,\boldsymbol{x}_n^H\boldsymbol{w}_n-\textstyle\frac{3}{4}\big(\boldsymbol{x}_n^H\boldsymbol{w}_n\big)^2\boldsymbol{w}_n^H\boldsymbol{x}_n -\textstyle\frac{1}{4}\big(\boldsymbol{w}_n^H\boldsymbol{x}_n\big)^3\Big)\boldsymbol{x}_n\Big]\\
&~~=R_m\,\EE\big[\boldsymbol{x}_n\boldsymbol{x}_n^H\big]\boldsymbol{w}_n -\textstyle\frac{3}{4}\EE\big[\boldsymbol{x}_n\boldsymbol{x}_n^H\boldsymbol{w}_n\boldsymbol{w}_n^H\boldsymbol{x}_n\boldsymbol{x}_n^H\boldsymbol{w}_n\big] -\textstyle\frac{1}{4}\EE\big[\big(\boldsymbol{w}_n^H\boldsymbol{x}_n\big)^3\boldsymbol{x}_n\big],
\end{split}
\end{equation}
We can show that\footnote{In (\ref{Eqstat2}), $\otimes$ denotes Kronecker product where each element of $(\boldsymbol{A}\otimes\boldsymbol{B})\in\mathbb{C}^{mp\times nq}$ is the product of an element of $\boldsymbol{A}\in\mathbb{C}^{m\times n}$ and an
element of $\boldsymbol{B}\in\mathbb{C}^{p\times q}$; the element in the $[p(i-1)+r]$th row
and $[q(j-1)+s]$th column of $\boldsymbol{A}\otimes\boldsymbol{B}$ is the $rs$th element $a_{ij} b_{rs}$ of $a_{ij}\boldsymbol{B}$ \cite{Harvillebook1997}; $\textrm{vec}[\boldsymbol{A}]$ is vector-valued function which assigns a (column-vector) value to $\boldsymbol{A}$ such that the $ij$th element of $\boldsymbol{A}$ is the $[(j-1)m+i]$th element of $\textrm{vec}[\boldsymbol{A}]$ \cite{Harvillebook1997}, and the $\textrm{mat}[\boldsymbol{a}]$ is a reverse operation which converts an $N^2 \times 1$ vector
$\boldsymbol{a}$ back to an $N\times N$ square matrix form \cite{GiHong2012}.}
\begin{equation}
\label{Eqstat2}
\begin{split}
&\EE\big[\boldsymbol{x}_n\boldsymbol{x}_n^H \boldsymbol{w}_n\boldsymbol{w}_n^H \boldsymbol{x}_n\boldsymbol{x}_n^H\boldsymbol{w}_n\big]=\EE\big[\boldsymbol{x}_n\boldsymbol{x}_n^H \boldsymbol{w}_n\boldsymbol{w}_n^H \boldsymbol{x}_n\boldsymbol{x}_n^H\big]\boldsymbol{w}_n \\ &~~~=\EE\big[\textrm{mat}\big[\textrm{vec}\big[\boldsymbol{x}_n\boldsymbol{x}_n^H \boldsymbol{w}_n\boldsymbol{w}_n^H \boldsymbol{x}_n\boldsymbol{x}_n^H\big]\big]\big]\boldsymbol{w}_n\\
&~~~=\EE\big[\textrm{mat}\big[\big((\boldsymbol{x}_n\boldsymbol{x}_n^H)^T\otimes(\boldsymbol{x}_n\boldsymbol{x}_n^H)\big)\textrm{vec}\big[\boldsymbol{W}_n\big]\big]\big]\boldsymbol{w}_n\\
&~~~=\textrm{mat}\big[\EE\big[(\boldsymbol{x}_n\boldsymbol{x}_n^H)^T\otimes(\boldsymbol{x}_n\boldsymbol{x}_n^H)\big]\textrm{vec}\big[\boldsymbol{w}_n\boldsymbol{w}_n^H\big]\big]\boldsymbol{w}_n,
\end{split}
\end{equation}
The matrix operation, \textrm{mat}[$\cdot$], as used in (\ref{Eqstat2}), however, is not an orthodox procedure, and is not supported necessarily by traditional digital signal processors. To resolve this,
alternatively, we may obtain a more elegant expression as follows:
\begin{equation}\begin{split}
&\EE\big[(\boldsymbol{x}_n^H\boldsymbol{w}_n)^2\boldsymbol{w}_n^H\boldsymbol{x}_n\boldsymbol{x}_n\big]=\EE\big[\boldsymbol{x}_n^H\boldsymbol{w}_n\boldsymbol{x}_n^H\boldsymbol{w}_n\boldsymbol{w}_n^H\boldsymbol{x}_n\boldsymbol{x}_n\big]\\
&~~=\EE\big[\boldsymbol{x}_n\boldsymbol{x}_n^H\boldsymbol{w}_n\boldsymbol{x}_n^H\boldsymbol{w}_n\boldsymbol{x}_n^T\boldsymbol{w}_n^\ast\big]\\
&~~=\EE\Big[\boldsymbol{x}_n\,\textrm{vec}\Big[\boldsymbol{x}_n\,\textrm{vec}\big[\boldsymbol{x}_n\boldsymbol{x}_n^H\big]^T\Big]^H\Big]\textrm{vec}\Big[\boldsymbol{w}_n\,\textrm{vec}\big[\boldsymbol{w}_n\boldsymbol{w}_n^H\big]^T\Big]
\end{split}
\end{equation}
Further, one may obtain:
\begin{equation}\begin{split}
&\EE\big[(\boldsymbol{w}_n^H\boldsymbol{x}_n)^3\boldsymbol{x}_n\big]=\EE\big[\textrm{vec}\big[(\boldsymbol{w}_n^H\boldsymbol{x}_n)^3\boldsymbol{x}_n\big]\big]\\
&~~~=\EE\big[\textrm{vec}\big[\boldsymbol{x}_n\boldsymbol{w}_n^H\boldsymbol{x}_n\boldsymbol{w}_n^H\boldsymbol{x}_n\boldsymbol{w}_n^H\boldsymbol{x}_n\big]\big]\\
&~~~=\EE\Big[\big(\boldsymbol{x}_{n}^{T}\otimes\boldsymbol{x}_{n}\big)\big(\boldsymbol{w}_{n}^\ast\otimes\boldsymbol{w}_{n}^H\big)\big(\boldsymbol{x}_{n}^{T}\otimes\boldsymbol{x}_{n}\big)\textrm{vec}\big[\boldsymbol{w}_n^H\big]\Big]\\
&~~~=\EE\Big[\big(\boldsymbol{x}_{n}^{T}\otimes\boldsymbol{x}_{n}\big)\big(\boldsymbol{w}_{n}^\ast\otimes\boldsymbol{w}_{n}^H\big)\big(\boldsymbol{x}_{n}^{T}\otimes\boldsymbol{x}_{n}\big)\Big]\boldsymbol{w}_n^\ast
\end{split}
\end{equation}
However, computing a statistics of $\boldsymbol{x}_n$ involving $\boldsymbol{w}_n$ is inadmissible. One of the feasible solutions is to evaluate:
\begin{equation}\begin{split}
&\EE\big[(\boldsymbol{w}_n^H\boldsymbol{x}_n)^3\boldsymbol{x}_n\big]=\EE\big[\boldsymbol{x}_n(\boldsymbol{w}_n^H\boldsymbol{x}_n\boldsymbol{w}_n^H\boldsymbol{x}_n\boldsymbol{w}_n^H\boldsymbol{x}_n)\big]\\
&~~=\EE\Big[\boldsymbol{x}_n\,\textrm{vec}\Big[\boldsymbol{x}_n\,\textrm{vec}\big[\boldsymbol{x}_n\boldsymbol{x}_n^T\big]^T\Big]^T\Big]\,\textrm{vec}\Big[\boldsymbol{w}_n\,\textrm{vec}\big[\boldsymbol{w}_n\boldsymbol{w}_n^T\big]^T\Big]^\ast
\end{split}
\end{equation}

Next, we can estimate required statistics either by taking ensemble average over a batch of data or iteratively updating the estimate at each time index. At index $n$, an iterative estimate of expectation $\EE[\boldsymbol{f}_n]$, where $\boldsymbol{f}_n$ is some matrix with random variable's entities, may be obtained as  \(\boldsymbol{S}_n=(1-\lambda)\boldsymbol{S}_{n-1}+\lambda\boldsymbol{f}_n\), \(0<\lambda<1\).
Next, using \(\boldsymbol{S}_n^{I}\), \(\boldsymbol{S}_n^{II}\), and \(\boldsymbol{S}_n^{III}\) to denote iterative estimates of  $\EE\big[\boldsymbol{X}_n\big]=\EE[\boldsymbol{x}_n\boldsymbol{x}_n^H]$, $\EE\Big[\boldsymbol{x}_n\,\textrm{vec}\big[\boldsymbol{x}_n\,\textrm{vec}\big[\boldsymbol{x}_n\boldsymbol{x}_n^H\big]^T\big]^H\Big]$, and $\EE\Big[\boldsymbol{x}_n\,\textrm{vec}\big[\boldsymbol{x}_n\,\textrm{vec}\big[\boldsymbol{x}_n\boldsymbol{x}_n^T\big]^T\big]^T\Big]$, respectively, we obtain feedforward
steepest descent \textsf{MMA2-2} (\textsf{SD-MMA2-2}) as given by:
\begin{subequations}\label{EqProposedStat}
\begin{empheq}[box=\fbox]{align}
& \nonumber \fbox{\textsf{SD-MMA2-2}} \\
& \nonumber\boldsymbol{w}_{n+1} =
 \boldsymbol{w}_n + \mu R_m\,\boldsymbol{S}_n^{I}\,\boldsymbol{w}_n\\
 &\nonumber\qquad\qquad\quad\,-\,\textstyle\frac{3}{4}\mu\,\boldsymbol{S}_n^{II}\,\textrm{vec}\Big[\boldsymbol{w}_n\,\textrm{vec}\big[\boldsymbol{w}_n\boldsymbol{w}_n^H\big]^T\Big]\\
 &\qquad\qquad\quad\,-\,\textstyle\frac{1}{4}\mu\,\boldsymbol{S}_n^{III}\,\textrm{vec}\Big[\boldsymbol{w}_n\,\textrm{vec}\big[\boldsymbol{w}_n\boldsymbol{w}_n^T\big]^T\Big]^\ast,\\
%
& ~\boldsymbol{S}_n^{I} = (1-\lambda)  \boldsymbol{S}_{n-1}^{I}+\lambda\,\boldsymbol{x}_n\,\boldsymbol{x}_n^H,\\
& ~\boldsymbol{S}_n^{II} = (1-\lambda) \boldsymbol{S}_{n-1}^{II}+\lambda\,
\boldsymbol{x}_n\,\textrm{vec}\Big[\boldsymbol{x}_n\,\textrm{vec}\big[\boldsymbol{x}_n\boldsymbol{x}_n^H\big]^T\Big]^H,\\
& ~\boldsymbol{S}_n^{III} = (1-\lambda) \boldsymbol{S}_{n-1}^{III}+\lambda\,
\boldsymbol{x}_n\,\textrm{vec}\Big[\boldsymbol{x}_n\,\textrm{vec}\big[\boldsymbol{x}_n\boldsymbol{x}_n^T\big]^T\Big]^T.
\end{empheq}\end{subequations}

Considering a fixed channel, assume that the (steady-state) estimates of statistics $\boldsymbol{S}_n^{I}$, $\boldsymbol{S}_n^{II}$ and $\boldsymbol{S}_n^{III}$ are available, say from the received large batch of data. Now, solving \(\partial J/\partial \boldsymbol{w}^\ast=\boldsymbol{0}\) and exploiting these available statistics, we obtain the following offline fixed-point steepest descent algorithm:
\begin{equation}
\label{EqProposedAlgoOffline}
\begin{split}
 \boldsymbol{w} \longleftarrow \frac{\Big[\,\overline{\boldsymbol{S}^{I}}\,\Big]^{-1}}{4R_m}\!\bigg( 3\,\overline{\boldsymbol{S}^{II}}\,\textrm{vec}\Big[\boldsymbol{w}\,\textrm{vec}\big[\boldsymbol{w}\boldsymbol{w}^H\big]^T\Big] ~~~~~~~~~~~~& \\
 +\,\overline{\boldsymbol{S}^{III}}\,\textrm{vec}\Big[\boldsymbol{w}\,\textrm{vec}\big[\boldsymbol{w}\boldsymbol{w}^T\big]^T\Big]^\ast\bigg)&
\end{split}\end{equation}
where $\overline{\boldsymbol{S}^{I}}$, $\overline{\boldsymbol{S}^{II}}$ and $\overline{\boldsymbol{S}^{III}}$ are offline estimates of $\boldsymbol{S}_n^{I}$, $\boldsymbol{S}_n^{II}$ and $\boldsymbol{S}_n^{III}$, respectively. However, note that the iteration (\ref{EqProposedAlgoOffline}) is found to be diverging which is a common problem in fixed-point procedure when matrix inverse is involved; see \cite[eq. (21) and details therein]{hyvarinen1999fast}. To improve this situation, we add a step-size in (\ref{EqProposedAlgoOffline}), obtaining
a stabilized (offline) fixed-point algorithm:
\begin{empheq}[box=\fbox]{align}
\label{EqProposedOfflineStable}
\begin{aligned}
& \fbox{\textsf{FP-MMA2-2}} \\
& \boldsymbol{w} \longleftarrow
\boldsymbol{w} + \mu\,\bigg( R_m\,\overline{\boldsymbol{S}^{I}}\,\boldsymbol{w}
-\textstyle\frac{3}{4}\overline{\boldsymbol{S}^{II}}\,\textrm{vec}\Big[\boldsymbol{w}\,\textrm{vec}\big[\boldsymbol{w}\boldsymbol{w}^H\big]^T\Big]
\\
& ~~~\qquad\qquad\qquad\qquad-\textstyle\frac{1}{4}\overline{\boldsymbol{S}^{III}}\,\textrm{vec}\Big[\boldsymbol{w}\,\textrm{vec}\big[\boldsymbol{w}\boldsymbol{w}^T\big]^T\Big]^\ast\bigg)
\end{aligned}\end{empheq}
where $\mu$ is step-size which may be made adaptive with iteration count. It is observed that a more certain convergence may be ensured if a $\mu$ much smaller than unity is selected (say, 0.1 or 0.01 for 4- or 16-\textsf{QAM}, respectively, with \(N=21\)). Here, we must mention that the evaluation of an optimal step-size for update (\ref{EqProposedOfflineStable}) is possible (see \cite{zarzoso2005blind,zarzoso2008optimal,zarzoso2010robust} for the idea), and has been left for future work.

\section{Simulation Results} We examine performance of proposed algorithm for the mitigation of interference caused by two Baud-spaced channels for 16-\textsf{QAM} signaling. The first channel, channel-1, is a voice-band telephone channel \(\boldsymbol{h}_n=[-0.005\) \( -\) \(0.004\textrm{j}, 0.009\) \( +\) \(0.03\textrm{j}, -0.024\) \( -\) \(0.104\textrm{j}, 0.854\) \( +\) \(0.52\textrm{j}, -0.218\) \(+\) \(0.273\textrm{j}, 0.049\) \(-\) \(0.074\textrm{j}, -0.016\) \(+\) \(0.02\textrm{j} ]\) taken from \cite{Picchi1987}. The second channel, channel-2, has a relatively large eigen-spread, and is given as
\(\boldsymbol{h}_n=[-0.023\) \( - \) \( 0.0345\textrm{j},
   0.0804\) \( - \) \( 0.0804\textrm{j},
   0.2068\) \( - \) \( 0.1149\textrm{j},
   0.678\) \( + \) \( 0.1378\textrm{j},
   0.1277\) \( + \) \( 0.0345\textrm{j},
  -0.1232\) \( - \) \( 0.1103\textrm{j}\),
  \(-0.023\) \( - \) \( 0.021\textrm{j},
   0.0176\) \( + \) \( 0.1196\textrm{j},
   0.0115\) \( + \) \( 0.0118\textrm{j} ]\). The signal-to-noise-ratio is $30$ dB. The equalizer length is $15$, initialized with a
unit spike at center tap, and all algorithms use step-size of $10^{-4}$.

The \textsf{ISI} measure in dB at $n$th time index is
\begin{equation}
\mathrm{\textsf{ISI}}_n= 10\log_{10}\left[\frac{1}{N_{\textrm{runs}}}\sum_{k=1}^{N_{\textrm{runs}}}
\frac{\sum_i |\boldsymbol{t}_{n,k}(i)|^2 - \max\{|\boldsymbol{t}_{n,k}|^2\}}
{\max\{|\boldsymbol{t}_{n,k}|^2\}}\right]
\end{equation}
where $\boldsymbol{t}_{n,k}$ is the overall channel-equalizer impulse response vector at index $n$ in the $k$th run of simulation. ${t}_{n,k}(i)$ represents the $i$th entity of
$\boldsymbol{t}_{n,k}$, and $\max\{|\boldsymbol{t}_{n,k}|^2\}$ represents the largest squared amplitude in $\boldsymbol{t}_{n,k}$.

\begin{figure}[htbp]
\centering
\begin{tabular}{l}
\hspace{-8mm}\includegraphics[scale=0.46]{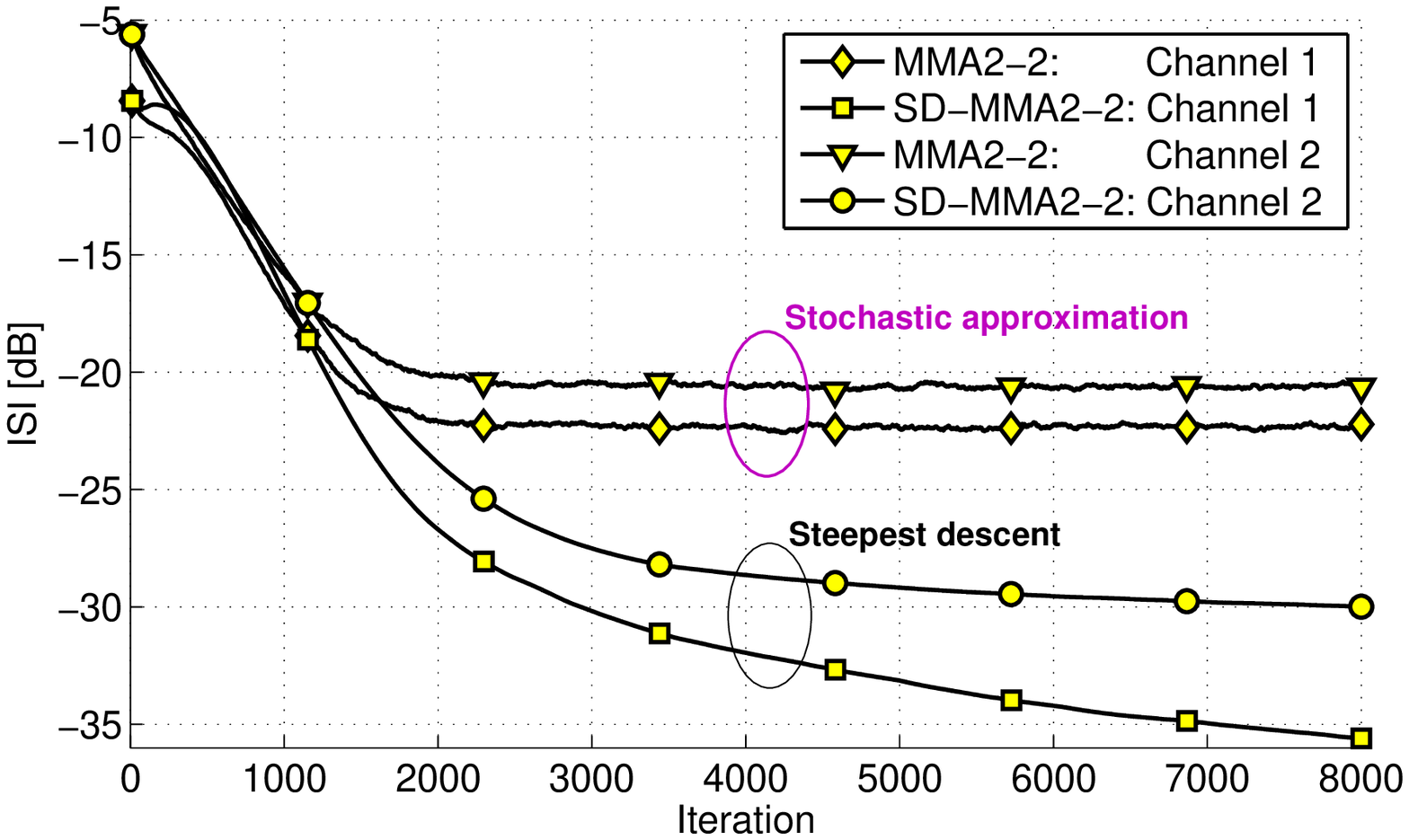}
(a)\\
\hspace{-8mm}\includegraphics[scale=0.46]{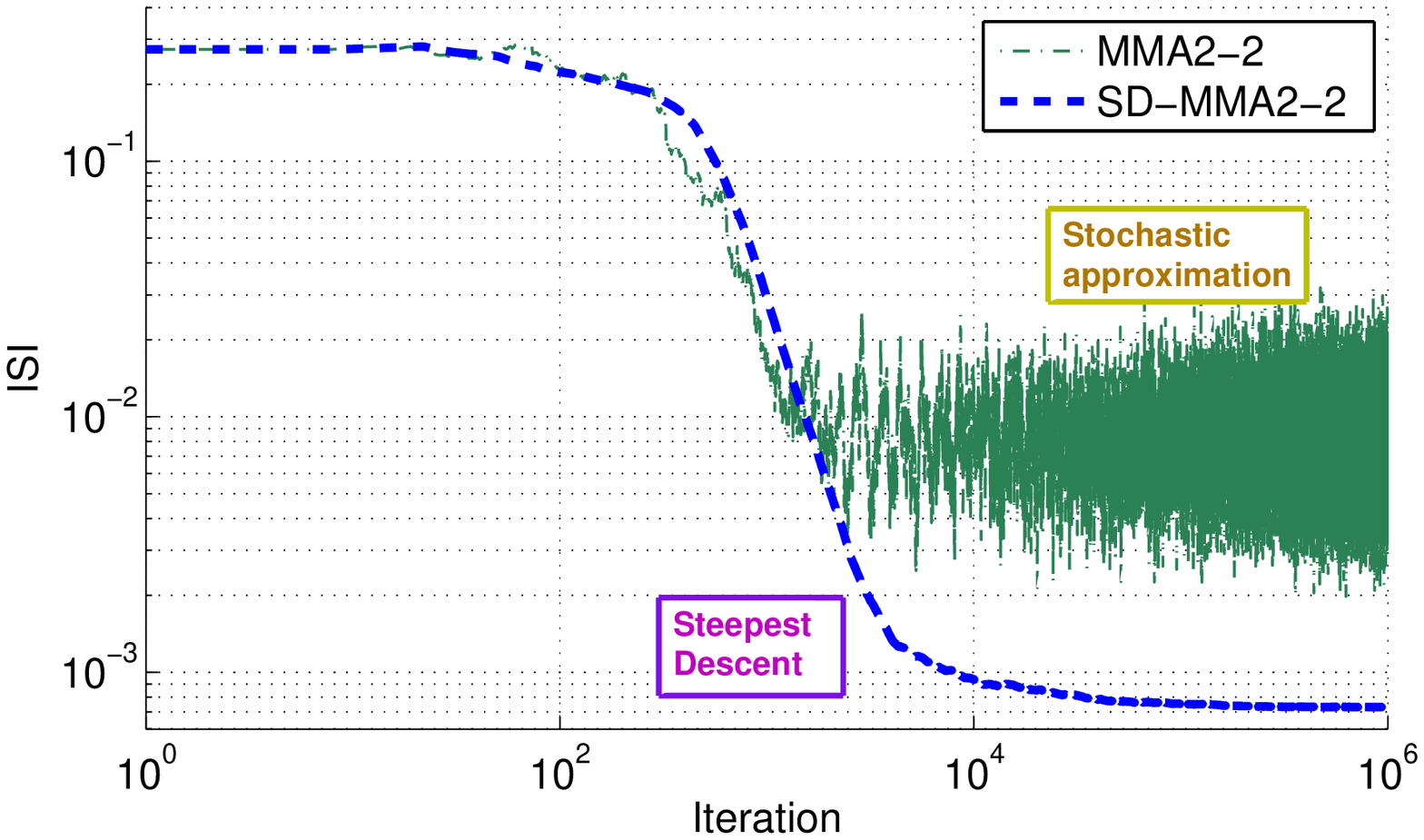}(b)
\end{tabular}
\caption{\textsf{ISI} convergence: (a) averaged trajectories. (b) Randomly selected single trajectories for channel-2.}\label{fig1}
\end{figure}

For fixed channels, we choose \(\lambda=1/n\) ($n$ is time index) so that the required statistics are estimated over all received data.
Fig. \ref{fig1}(a) demonstrates convergence behaviors of \textsf{MMA2-2} and \textsf{SD-MMA2-2}, averaged over $400$ and $50$ independent runs ($N_{\textrm{runs}}$), respectively. We notice that the \textsf{ISI} mitigation achieved by \textsf{SD-MMA2-2} is far better in steady-state when allowed to converge at the same rate as that of \textsf{MMA2-2}. In Fig.~\ref{fig1}(b), single trajectory of \textsf{ISI} convergence of each \textsf{MMA2-2} and \textsf{SD-MMA2-2} is shown. We can note that the \textsf{SD-MMA2-2} exhibits far smoother and more stable convergence than \textsf{MMA2-2} (for fixed channel scenario), and this is the reason why we used fewer independent runs for the ensemble averaging of \textsf{ISI} trajectories in \textsf{SD-MMA2-2} than \textsf{MMA2-2}.

\section{Conclusions}

A steepest descent implementation of \textsf{MMA2-2} for blind signal recovery has been proposed and demonstrated to mitigate \textsf{ISI}. The proposed equalizer has been found to yield better steady-state performance than stochastic approximate gradient descent \textsf{MMA2-2}. Thus, the proposed approach seems to be quite a promising substitute for traditional counterpart on fixed channels. Future work includes: (a) application to time-varying channels, (b) evaluation of optimal step-sizes, and (c) application to \textsf{MIMO} systems.

\begin{small}
\bibliographystyle{unsrt}
\bibliography{SteepestDescentMMA_Archiv}

\begin{thebibliography}{10}

\bibitem{abrar2010ieeetn}
Shafayat Abrar and Asoke~Kumar Nandi.
\newblock Blind equalization of square-{QAM} signals: a multimodulus approach.
\newblock {\em IEEE Transactions on Communications}, 58(6), 2010.

\bibitem{abrar2006ieeespl}
Shafayat Abrar and Syed~Ismail Shah.
\newblock New multimodulus blind equalization algorithm with relaxation.
\newblock {\em IEEE Signal Processing Letters}, 13(7):425--428, 2006.

\bibitem{azim2015}
Ali~Waqar Azim, Shafayat Abrar, Azzedine Zerguine, and Asoke~Kumar Nandi.
\newblock Steady-state performance of multimodulus blind equalizers.
\newblock {\em Signal Processing}, 108:509--520, 2015.

\bibitem{han2012thesis}
Huy-Dung Han.
\newblock {\em Batch algorithms for blind channel equalization and blind
  channel shortening using convex optimization}.
\newblock University of California, Davis, 2012.

\bibitem{han2017convex}
Huy-Dung Han, Zhi Ding, and Muhammad Zia.
\newblock A convex relaxation approach to higher-order statistical approaches
  to signal recovery.
\newblock {\em IEEE Transactions on Vehicular Technology}, 66(1):188--201,
  2017.

\bibitem{shah2015}
Syed Awais~Wahab Shah, Karim Abed-Meraim, and Tareq~Yousef Al-Naffouri.
\newblock Multi-modulus algorithms using hyperbolic and givens rotations for
  blind deconvolution of mimo systems.
\newblock In {\em IEEE International Conference on Acoustics, Speech and Signal
  Processing}, pages 2155--2159. IEEE, 2015.

\bibitem{daumont2010}
Steredenn Daumont and Daniel Le~Guennec.
\newblock An analytical multimodulus algorithm for blind demodulation in a
  time-varying {MIMO} channel context.
\newblock {\em International Journal of Digital Multimedia Broadcasting}, 2010.

\bibitem{Debals2016}
Otto Debals, Muhammad Sohail, and Lieven~De Lathauwer.
\newblock Analytical multi-modulus algorithms based on coupled canonical
  polyadic decompositions.
\newblock Technical report, Technical Report 16-150, ESAT-STADIUS, KU Leuven,
  Leuven, Belgium, 2016.

\bibitem{GiHong2012}
Huy-Dung Han and Zhi Ding.
\newblock Steepest descent algorithm implementation for multichannel blind
  signal recovery.
\newblock {\em IET Communications}, 6(18):3196--3203, 2012.

\bibitem{Harvillebook1997}
David~A Harville.
\newblock {\em Matrix algebra from a statistician's perspective}, volume~1.
\newblock Springer, 1997.

\bibitem{hyvarinen1999fast}
Aapo Hyvarinen.
\newblock Fast and robust fixed-point algorithms for independent component
  analysis.
\newblock {\em IEEE Transactions on Neural Networks}, 10(3):626--634, 1999.

\bibitem{zarzoso2005blind}
Vicente Zarzoso and Pierre Comon.
\newblock Blind and semi-blind equalization based on the constant power
  criterion.
\newblock {\em IEEE Transactions on Signal Processing}, 53(11):4363--4375,
  2005.

\bibitem{zarzoso2008optimal}
Vicente Zarzoso and Pierre Comon.
\newblock Optimal step-size constant modulus algorithm.
\newblock {\em IEEE Transactions on communications}, 56(1), 2008.

\bibitem{zarzoso2010robust}
Vicente Zarzoso and Pierre Comon.
\newblock Robust independent component analysis by iterative maximization of
  the kurtosis contrast with algebraic optimal step size.
\newblock {\em IEEE Transactions on Neural Networks}, 21(2):248--261, 2010.

\bibitem{Picchi1987}
Giorgio Picchi and Giancarlo Prati.
\newblock Blind equalization and carrier recovery using a ``stop-and-go''
  decision-directed algorithm.
\newblock {\em IEEE Transactions on Communications}, 35(9):877--887, 1987.

\end{thebibliography}
\end{small}

\end{document}